\documentclass{pazh_engl}
\usepackage{epsf}
\usepackage{flushrt}
\usepackage{graphicx}
\usepackage{pstricks}

\def\SAX{\mbox{SAX J2103.5+4545}}
\def\deg{$^\circ$}
\def\etal{{\it et~al.}}
\def\arcsec{$^{\prime\prime\,}$}

\begin{document}

\sloppypar

\title{\bf First results of observations of transient pulsar 
SAXJ2103.5+4545 with the INTEGRAL observatory.}

\author{\copyright 2003 A.Lutovinov\inst{1,2}, S.Molkov\inst{1}, M.Revnivtsev\inst{2,1}}

\institute{Space Research Institute, Moscow, Russia
\and
Max-Plank Institute f\"ur Astrophysik, Garching, Germany}

\authorrunning{LUTOVINOV}

\date{8 May 2003}

\abstract{We present preliminary results of observations of X-ray pulsar
SAX J2103.5+4545 with INTEGRAL observatory in Dec 2002. Maps of this 
sky region in energy bands 3-10, 15-40, 40-100 and 100-200 keV are presented. 
The source is significantly detected up to energies of $\sim100$ keV. 
The hard X-ray flux in the 15-100 energy band is variable, that could be
connected with the orbital phase of the binary system. 
We roughly reconstructed the source spectrum using its comparison 
to that of Crab nebula. It is shown that the parameters of the source
spectrum in 18-150 keV energy range are compatible with that obtained earlier by RXTE observatory}

\titlerunning{Observations of pulsar \SAX\ with the INTEGRAL observatory.}

\maketitle

\section*{Introduction}

The transient X-ray pulsar \SAX\ was discovered by the BeppoSAX observatory 
during its outburst in 1997  (Hulleman \etal. 1998). The source observations 
with the RXTE satellite during its next outburst in 1999 have shown that
along with $\sim 358$s pulsations the source demonstrates significant
periodicity caused by a binary orbit. Based on these
observations Baykal \etal (2000) showed that the pulsar is a member of a 
binary system with an eccentric orbit  
($e \approx 0.4$) and orbital period of $\sim 12.68$ days.

The source spectrum was described a power law with a photon index of
$\sim1.3$ and high energy cutoff (Hulleman \etal\ 1998, Baykal \etal\ 2002),
that is typical for accreting X-ray pulsars. In addition, at the low 
energies the spectrum was  
modified by an absorption with a hydrogen column density of 
$N_H=3.8\times10^{22}$ atoms cm$^{-2}$. In addition, a fluorescent 
line of neutral iron at $\sim6.4$ keV was detected in the spectrum.

Analysis of the source light curve have shown that its intensity strongly
 variates over the orbital cycle and reaches the maximum near the periastron 
(Baykal \etal\ 2000). Such dependence of intensity on an orbital phase 
is quite typical for 
the high-eccentric binary systems with massive companions -- stars of early
spectral classes $O-B$. It was assumed that B-type star 
HD200709 is a possible companion of the pulsar (Hulleman \etal\ 1998),
however, it is still not confirmed.

It is necessary to note that the pulsar have demonstrated high stability
of the shape of its spectrum over years of observations
(Baykal \etal 2002), that gives us a possibility to compare the RXTE
results with results obtained by other observatories, in particular INTEGRAL.

According to results of observation of all-sky monitor (ASM) of the RXTE 
observatory the source remains relatively bright during last  
years and demonstrates the flaring activity. The off-flare part 
of its emission is rising during last several months.  

This work is dedicated to an investigation of a hard emission from
pulsar \SAX\ using the public data of the INTEGRAL observatory.

\section*{Observations and Data Analysis}

\subsection*{INTEGRAL}

The international gamma-ray observatory INTEGRAL was successively launched 
to its orbit with Russian rocket PROTON from the cosmodrom Baikonur on
Oct 17, 2002. It consists of instruments designed for investigation of cosmic
sources in a wide energy band 3-10000 keV: 1) the IBIS telescope, 
consisting of two detectors ISGRI and PICsIT working in the energy band 
15-10000 keV, and allowing to localize sources
with accuracy down to 30\arcsec;
2) the spectrometer SPI, working in the energy band 15-8000 keV and designed
for an accurate spectroscopy with the energy resolution of 
$E/\delta E\simeq 500$ (at 1 MeV); 3) the X-ray monitor JEM-X 
with effective energy range 3-35 keV. All telescopes are using the principle 
of coding aperture. The fields of view (FOV) of IBIS and JEM-X telescope
whose data were used in this work are: for the IBIS -- 25\deg$\times$25\deg\ 
full field of view, 9\deg$\times$9\deg\ fully coded FOV (FCFOV); for the JEM-X
 -- 4.8\deg\ in a diameter FCFOV. More detailed descriptions of the 
observatory, instruments, organizations of observations etc could be found in 
 Winkler (1996, 1999).

\begin{figure*}[]

\hbox{
\includegraphics[width=\columnwidth]{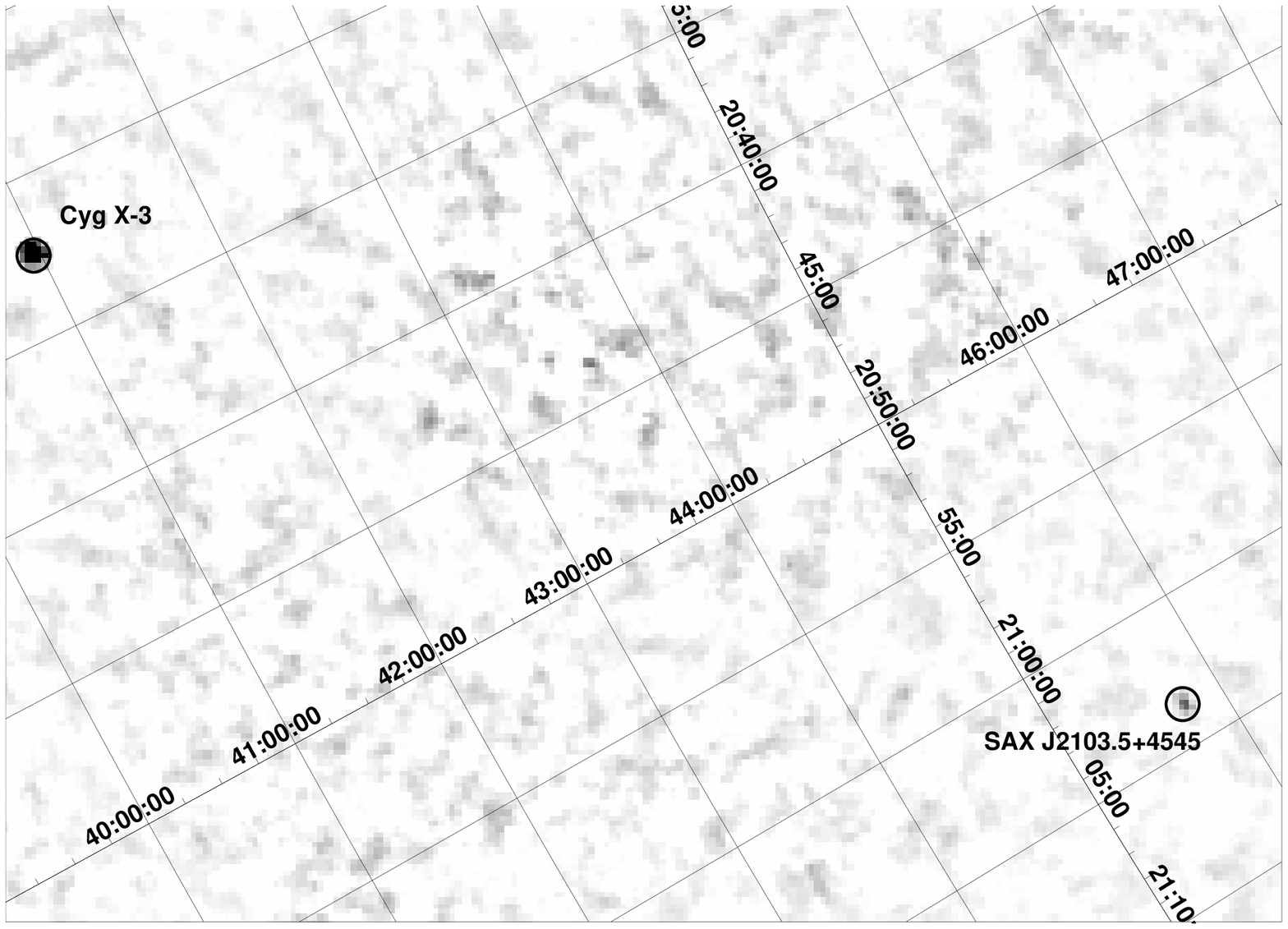}
\includegraphics[width=\columnwidth]{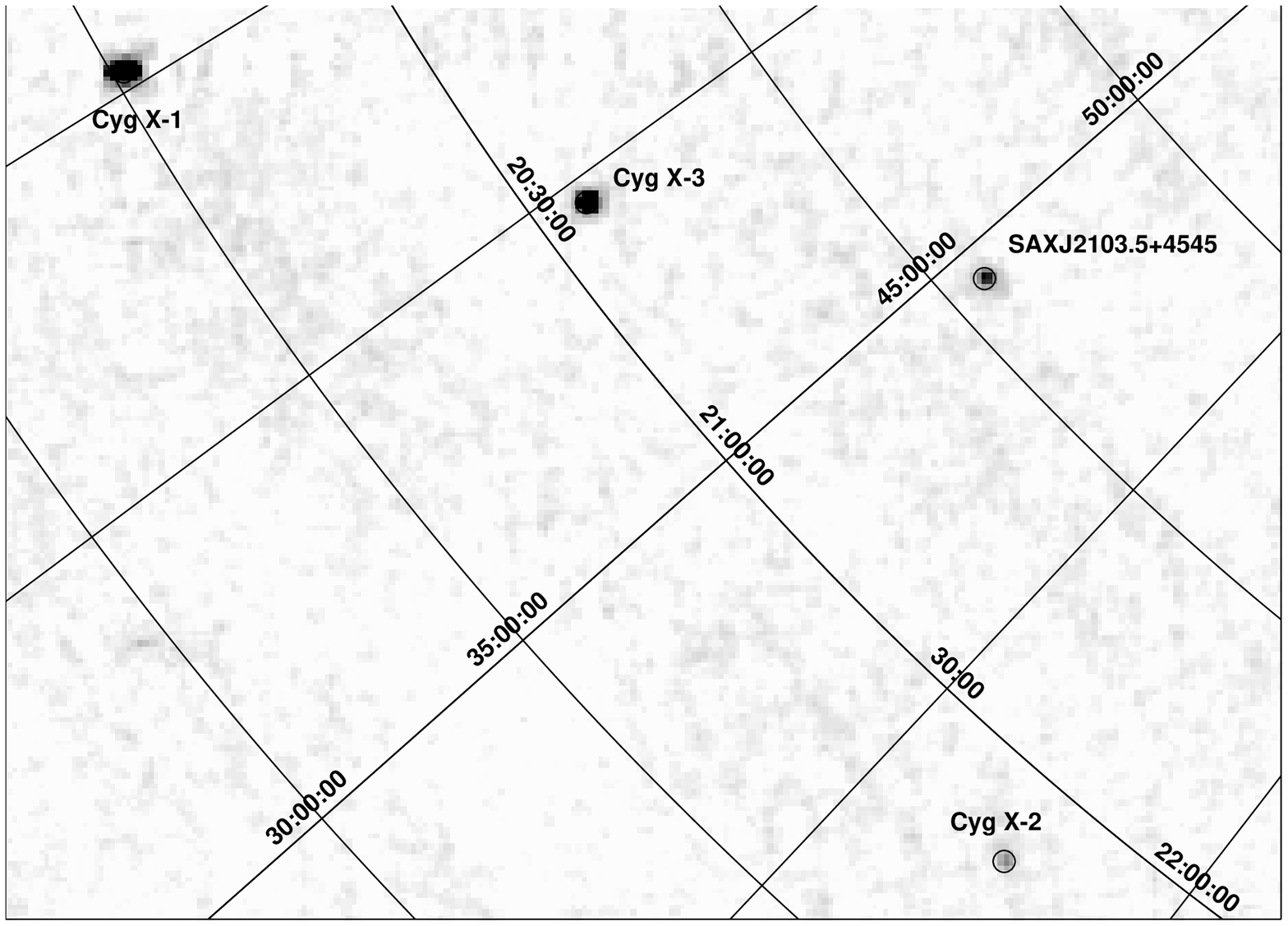}
}

\vspace{2mm}
\hbox{
\includegraphics[width=\columnwidth]{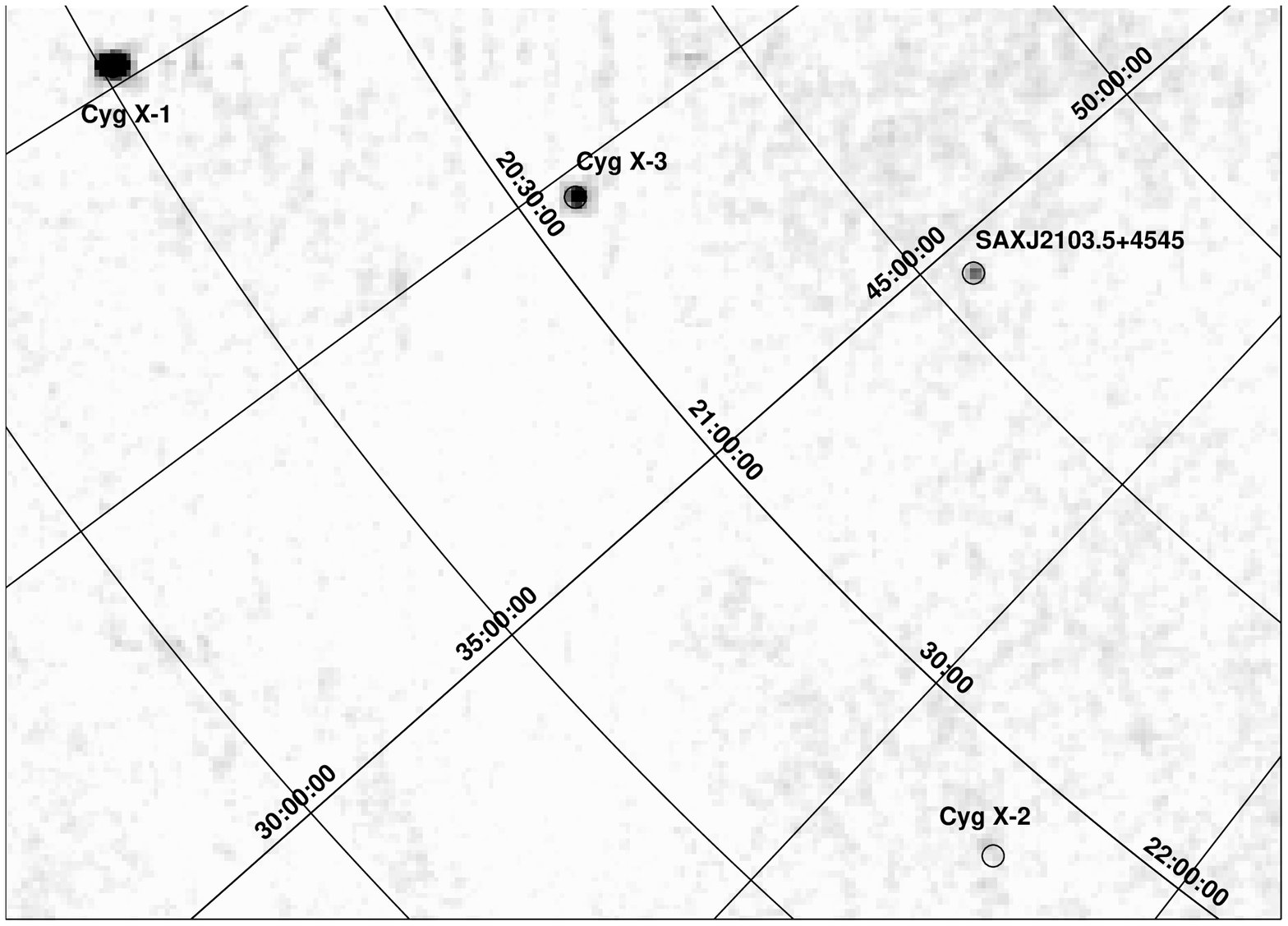}
\includegraphics[width=\columnwidth]{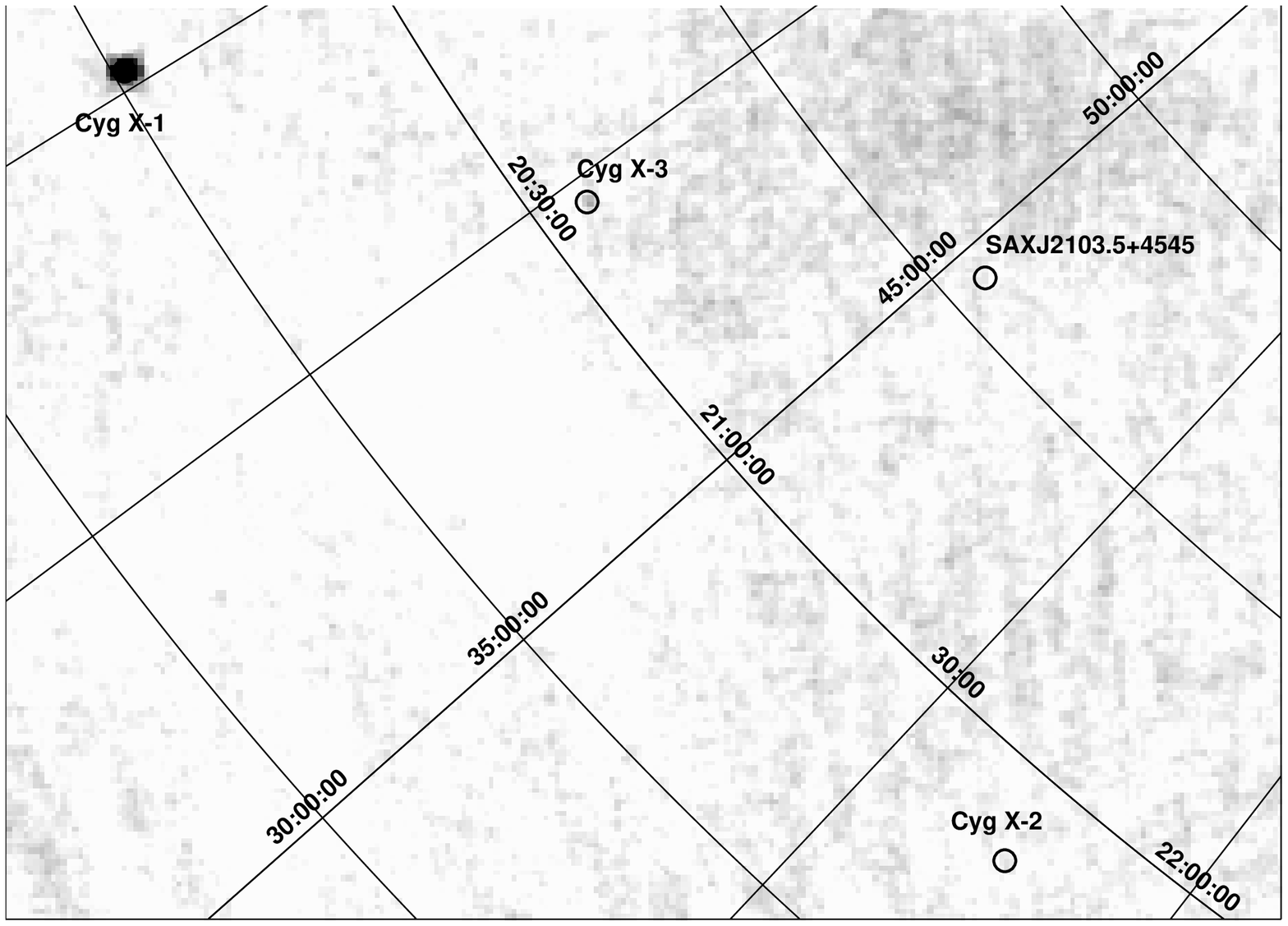}
}
\caption{Maps of the sky around SAX J2103.5+4545, obtained with the 
help of INTEGRAL telescopes on December 2002. The image obtained by the JEM-X 
monitor in the 3-10 keV band is presented at the left-upper panel. At the 
right-upper, left-down and right-down panels we present images obtained by 
the IBIS
telescope in 15-40, 40-100 and 100-200 energy bands, respectively. 
\label{image}}
\end{figure*}

In this work we used the public data of the calibration observations 
of Cyg X-1 sky region, obtained by the INTEGRAL observatory in December 
of 2002 (23 and 25 revolutions, intervals 
UT 2002-12-21 09:14:29 -- 2002-12-23 20:41:11 and UT 2002-12-28 11:53:17
 -- 2002-12-29 21:42:10, respectively).

Calibrations of instruments of observatory is still not
finished, and right now it is not possible for us to make full 
analysis of data of all 
telescopes. Therefore the main attention in our paper
would be paid to the data of IBIS telescope which have the 
largest effective area. We also use the data, obtained by the JEM-X monitor 
for the image reconstruction of the sky in soft energy band (3-10 keV). 

For the data reduction we used standard programs of package IDAS 1.0,
 distributed by INTEGRAL Science Data Center
(ISDC, http://isdc.unige.ch). Version 1.0 of the IDAS package doesn't allow
us to make full spectral and timing analysis of data, therefore 
for the preliminary spectral analysis of pulsar \SAX\ we used 
the big set of public calibration observations of Crab nebula obtained 
in February of 2003 (revolutions 39--45). Taking into account the present
uncertainties of the calibrations that lead to
strong dependence of the reconstructed source intensity on its distance 
from the center 
of the field of view (obtained from Crab nebula analysis), especially 
in low channels of the ISGRI detector, in our subsequent analysis
we used only data of observations
when \SAX\ was inside the FCFOV, where the intensity of the source should be
relatively stable. Analysis of the Crab nebulae gave us the count rate 
150 counts/s and 147 counts/s in 15-40 and 40-100 keV energy bands, 
respectively.

Reconstruction of the sky regions was made using whole set of data
(not only FCFOV data).
Thus the total exposures of ISGRI detector were $\sim260$ ksec for the image 
reconstruction and $\sim55$ ksec for spectral 
and timing analysis. Note that due to smaller field of view of the 
JEM-X monitor its full (not only FCFOV) exposure time was close to the latter value.

\subsection*{RXTE}

For the comparison of spectra obtained with INTEGRAL with previous
measurements we have analyzed the data of RXTE satellite 
(Bradt, Rothschild \& Swank 1993), that numerously observed the source
during period 1997-2002. Besides, in order to obtained the information
 about the sources behavior at the energy band lower than 15 keV 
we have used the data of all sky monitor (ASM) of RXTE 
(http://xte.mit.edu/ASM\_lc.html).

Main instruments of RXTE observatory are spectrometers PCA and HEXTE, 
covering together wide energy band 3-250 keV. Spectrometer PCA
consists of 5 independent gas proportional counters. Field of view
of the spectrometer is limited by 1$^\circ$ collimator. Effective energy band
of the PCA is 3-20 keV,  effective area $\sim6400$ sq.cm at eneries 6-7 keV, 
energy resolution $\sim18$\%. Spectrometer HEXTE consists of two clusters
of 4 detectors NiI(Tl)/CsI(Na), rocking at 16 sec time scale for background
measurements. At the any certain time only one HEXTE cluster is
 observing the sources, therefore the effective area of the instrument 
is $\sim$700 sq.cm. Bandpass of the instrument is 15-250 keV.

For the spectral extraction we have used the data of $\sim$190 observations, 
performed in 1997-2001. Effective exposure of used observations
is $\sim$560 ksec. As it was shown the spectral shape of the source
is relatively stable (Baykal et al. 2002), therefore for spectral analysis
we have averaged data of all observations.

For data reduction of the RXTE observatory we have used standard
programs of FTOOLS 5.2 package.

\section*{Results}

\subsection*{Sky maps}

The maps of the sky region around the pulsar \SAX\, obtained by the 
JEM-X telescope and ISGRI detector of IBIS telescope are shown in Fig.\ref{image} in
3-10, 15-40, 40-100 and 100-200 keV energy bands. From these images
it is seen that the source is statistically significantly detected 
up to energies of $\sim100$ keV. Along with \SAX\ three known sources were 
detected in the IBIS field of view -- a black-hole candidate Cyg X-1,
low-mass X-ray binary Cyg X-2 and binary system Cyg X-3 (last was
detected also by JEM-X). Note that Cyg X-1 and Cyg X-3 are still
bright and are clearly seen in a hardest energy band 100-200 keV.

\subsection*{Light curves}

\begin{figure}[]
\includegraphics[width=\columnwidth,clip]{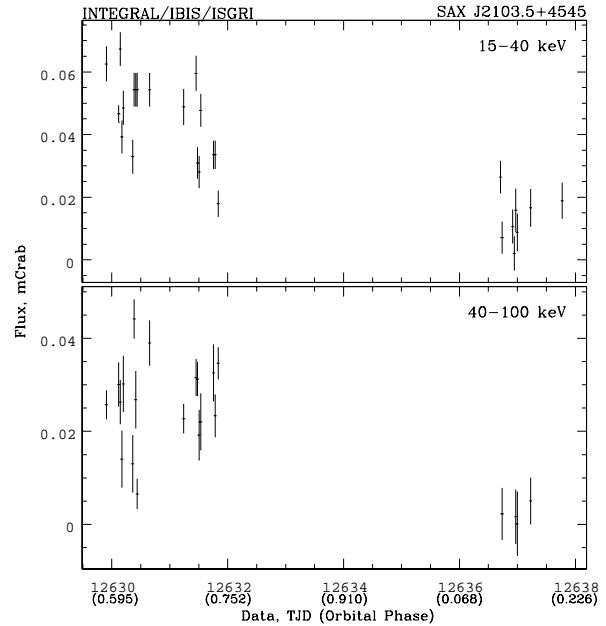}
\caption{Light curves of \SAX, obtained by the ISGRI detector of 
IBIS telescope in 15-40 
and 40-100 keV energy bands. Each point corresponds to
one telescope pointing observations with exposure of 
$\sim2200$ sec. \label{lcurve}}
\end{figure}

As it was mentioned in the Introduction the intensity of the pulsar emission
depends on its orbital phase and reaches its maximum near the periastron 
passage
(Baykal\etal\ 2000). The INTEGRAL observations show similar behavior
in hard X-ray and gamma-ray energy bands. Light curves 
of the pulsar obtained by IBIS in 15-40 and 40-100 keV energy bands 
are shown in Fig.\ref{lcurve}. For easier comparison of this picture with 
result obtained by Baykal \etal\ (2000), orbital phases are also
indicated on the Time axis (the orbital parameters were taken from the same
work). The flux detected from the system in both energy bands is highest 
near the orbital phase 0.5 and strongly decreases to the phases of 
0.1-0.2. Unfortunately due to absence of data between Dec. 23 and 28, 2002,
connected with the INTEGRAL calibration observations of empty fields,
we can not build the pulsar light curve during whole orbital cycle.

\subsection*{Spectrum}

In order to describe the spectrum of the source we used a ratio of fluxes 
measured by IBIS/ISGRI detector in different energy bands to the fluxes 
of Crab nebula measured by the ISGRI detector in the same energy band. 
The analysis of a set of observations of the Crab have shown that 
such method of investigation of sources allows us to roughly estimate
their spectral shape, but due to imperfection of first instrument
calibrations the systematic uncertainties play big role.
The observed amplitude of such uncertainties of the measurements
of sources fluxes can reach 10-20\% in different energy bands. 
It is important to note that as it was mentioned above we used 
only data of observations when both \SAX\ and Crab pulsar
were in the FCFOV.

\begin{figure} [t]
\includegraphics[width=\columnwidth,bb=35 190 565 720,clip]{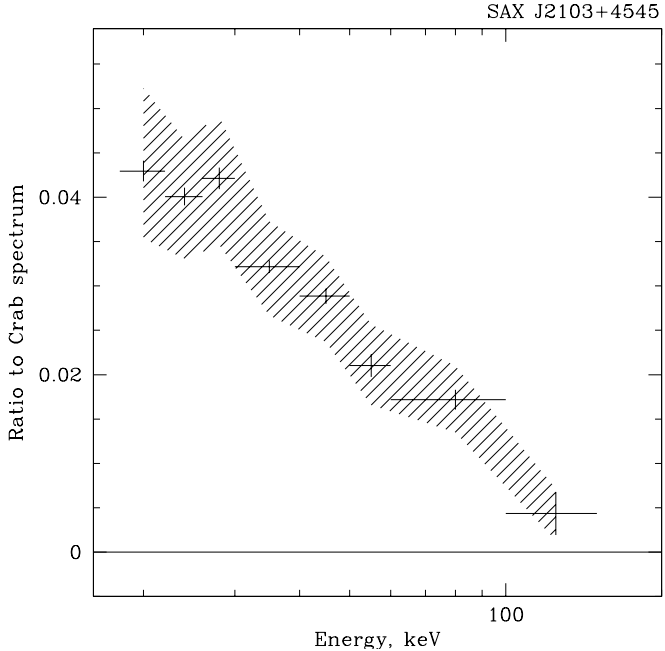}
\caption{Ratio of spectrum of \SAX\ to the spectrum of Crab  (IBIS/ISGRI data).
The region of possible uncertainties (statistic and systematic) of 
obtained values is shown by shadow.\label{rat_sp}}
\end{figure}

Using the known shape of spectrum of the Crab we can calculate 
the flux from \SAX\ in the units of phot/cm$^2$/s in each 
energy band. After this we can fit the obtained spectrum by different models. 
The ratio of the \SAX\ spectrum to the Crab spectrum is shown in
Fig.\ref{rat_sp}. Note that the procedure of the spectral reconstruction,
described above, is quite complex, imperfect and subject to different 
uncertainties. Therefore the $\chi^2$ values obtained during the 
fitting procedure can not be treated as a correct measurement of 
the quality of applied models.

For the approximation of the source spectrum we used two simplest models -- 
a power law  ($dN/dE \propto E^{-\Gamma}$) and the power law with high 
energy cutoff 
($dN/dE\propto E^{-\Gamma} \exp(-(E-E_{\rm cut})/E_{\rm fold}) $),
which is a typical model for approximation of spectra of accreting X-ray 
pulsars. The latter model was successively used for the describing
the source spectrum in a wide energy range obtained during its outbursts 
in 1997 and 1999. (Baykal \etal\ 2002)

\begin{table}[b]
\caption{Best-fit parameters of the \SAX\ spectrum approximation. IBIS/ISGRI data 
(18-150 keV)}
\begin{tabular}{l|c}
\hline
\multicolumn{2}{c}{Power law ($dN/dE\propto E^{-\Gamma}$)}\\
\hline
Photon index, $\Gamma$&$2.8\pm0.3$\\
Flux, 18-150 keV, $10^{-10}$ erg/cm$^2$/s&$6.2$\\
$\chi^2$(d.o.f)&12.2(6)\\
\hline
\multicolumn{2}{c}{$dN/dE\propto E^{-\Gamma}\exp(-(E-E_{\rm cut})/E_{\rm fold})$}\\
\hline
Photon index,$\Gamma$&1.3$^a$\\
$E_{\rm cut}$, keV&7.8$^a$\\
$E_{\rm fold}$, keV&$26\pm1$\\
Flux, 18-150 keV, $10^{-10}$ erg/cm$^2$/s&$5.7$\\
$\chi^2$(d.o.f)&5.2(6)\\
\hline
\end{tabular}

\vspace{2mm}

\begin{list}{}
\item $^a$ -- Because of absence of spectral information 
in the energy band
lower than 18 keV the parameters, related to this energy band, were fixed 
at values obtained by Baykal \etal\ 2002. 

\end{list}
\end{table}

\begin{figure}
\includegraphics[width=\columnwidth,bb=35 190 565 720,clip]{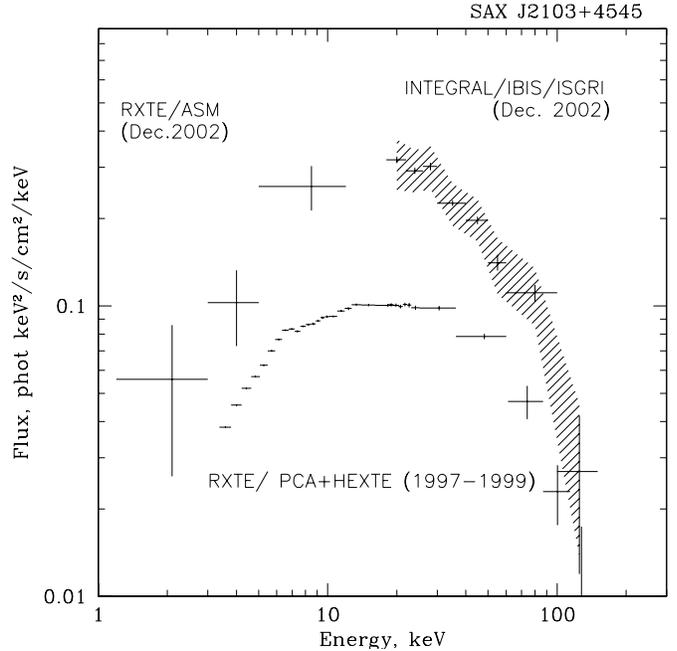}

\caption{Spectra of X-ray pulsar \SAX\ obtained by INTEGRAL and RXTE 
observatories (the normalization of the RXTE/PCA+HEXTE spectrum is arbitrary.  
The region of possible uncertainties (statistic and systematic) of 
obtained INTEGRAL/IBIS values is marked by shadow. At energies
lower than 15 keV measurements of RXTE/ASM are presented.
\label{spectr}}
\end{figure}

Best-fit parameters of approximations of the source spectrum
are presented in the Table 1. It is necessary to note that in the
spectral approximations we included the 15\% systematic uncertainties
to the fluxes measured in all energy bands.
Due to absence of information about source spectrum in the energy band 
lower that 18 keV we fixed several parameters at the values obtained 
by the RXTE observatory during the source bright state (Baykal \etal 2002).
The \SAX\ spectrum, obtained by the ISGRI detector of the IBIS telescope
is shown in Fig.\ref{spectr}.  Simultaneous measurements of the source flux
by RXTE/ASM in 1.3-3, 3--5, and 5-12 keV energy bands are also 
presented.

In order to compare the results of INTEGRAL observations with results 
of archived observations of \SAX\ in Fig. \ref{spectr} we also present 
the typical spectrum of the source in its bright state. For this
spectrum we used combined data of RXTE/PCA and RXTE/HEXTE spectrometers.
It is seen that the spectrum of \SAX\ obtained with IBIS telescope
is quite compatible with that obtained by RXTE observatory.

\section*{Conclusion}

The transient X-ray pulsar \SAX\ was observed by the INTEGRAL observatory 
during calibration observations of the Cyg X-1 field on December 2002. 
Analysis showed that the source is detected by IBIS telescope
up to energies $\sim100$ keV. The shape of the source spectrum agrees
with results of RXTE observatory.   
The spectrum of the pulsar in the energy range 18-150 keV can be described 
either by the simple power law with photon index of $\sim2.8$ or by the 
model of power law with high energy cutoff. 
The parameter $E_{fold}=26\pm1$ keV of this model is in a good agreement 
with results of RXTE observations  - $E_{fold}=27.1\pm0.9$ keV (Baykal \etal\ 
2002).  

We reconstructed the source light curves and showed that the source flux 
in the 15-100 keV energy band is variable that presumably could be
caused by a binary orbital variations. 
The maximum of detected hard X-ray flux (18-100 keV)
occurs nearly at periastron passage, that is in agreement with RXTE/ASM
results.

The increase of the the number of the source observations by inclusion
of data of INTEGRAL Galactic Plane Scans would allow us to cover the whole
orbital cycle, to increase our sensitivity at the energy band 100-200 keV
and to search for the cyclotron feature in the source spectrum, 
predicted at $\sim 140$ keV by  Baykal \etal\ (2002)

\bigskip
{\it
Research has made use of data obtained 
through the INTEGRAL Science Data Center (ISDC), Versoix, Switzerland, 
and  High Energy Astrophysics Science Archive Research Center 
Online Service, provided by the NASA/Goddard Space Flight Center.
Authors are thankful to the group of INTEGRAL Science Data Center
for their hospitality and computing resources.

\bigskip

This work was partially supported by grant of the RFBR 03-02-06772 and 
program of Russian Academy of Sciences ``Time-varying phenomena in 
astronomy''.

}

\section*{References}

\indent

Baykal A., Stark M., Swank J., Astrophys. J. {\bf 544}, L129
 (2000)

Baykal A., Stark M., Swank J., Astrophys. J. {\bf 569}, 903
 (2002)

Bradt, Rothschild, Swank, Astron. Astrophys. Suppl. Ser.  {\bf 97}, 355 (1993)

Winkler C., Astron. Astrophys. Suppl. {\bf 120}, 637 (1996)

Winkler C., Astrophys. Lett. \& Comm.  {\bf 39}, 309 (1999) 

Hulleman F., in't Zand J., Heise J., Astron. Astrophys. 
{\bf 337}, L25 (1998)

\end{document}